# Synthesis and systematic optical investigation of selective area droplet epitaxy of InAs/InP quantum dots assisted by block copolymer lithography

**Artem Shikin[1], Elizaveta Lebedkina[1], Czcibor Ciostek[2], Paweł Holewa[2], Sokol Ndoni[3], Kristoffer Almdal[3], Kresten Yvind[1], Marcin Syperek[2] and Elizaveta Semenova[1,*]**

[1] *DTU Fotonik, Technical University of Denmark, Kongens Lyngby DK-2800, Denmark*
[2] *Wroclaw University of Science and Technology, Wroclaw 50-370, Poland*
[3] *DTU Nanotech, Technical University of Denmark, Kongens Lyngby DK-2800, Denmark*
*\*esem@fotonik.dtu.dk*

**Abstract:** We report on the systematic investigation of the optical properties of a selectively grown quantum dot gain material assisted by block-copolymer lithography for potential applications in active optical devices operating in the wavelength range around 1.55 µm and above. We investigated a new type of diblock copolymer PS-b-PDMS (polystyrene-block-polydimethylsiloxane) for the fabrication of silicon oxycarbide hard mask for selective area epitaxy of InAs/InP quantum dots. An array of InAs/InP quantum dots was selectively grown via droplet epitaxy. Our detailed investigation of the quantum dot carrier dynamics in the 10-300 K temperature range indicates the presence of a density of states located within the InP bandgap in the vicinity of quantum dots. Those defects have a substantial impact on the optical properties of quantum dots.

## 1. Introduction

For more than 30 years quantum dots (QDs) have attracted significant attention due to their distinctive delta-like carrier density of states and the carrier dynamics [1] which open unique possibilities for device realizations. Self-assembled In(Ga)As QDs on GaAs substrates have, after several decades of research, finally fulfilled some of the early predictions for low dimensional semiconductor gain media [2]. The lowest threshold current densities [3] and lowest temperature dependence [4] have been demonstrated with QD lasers, making them superior to the more standard quantum well devices [5]. Furthermore, the dynamic properties of the QD gain medium [1,6] have allowed for notably improved temporal and noise [7] performances of lasers [8] and amplifiers, something that was theoretically predicted prior to the experimental demonstration [9]. Unfortunately, QD material on GaAs has a limited spectral coverage 0.9-1.31 µm. In order to achieve longer wavelength emission, the InP-based material system is mainly used. However, the QDs on this substrate have so far not been able to reproduce the superior properties of their short wavelength counterparts. Elongated "dash" like [10] QDs with a high in-plane aspect ratio and a broad size distribution are typically obtained [11]. Although laser devices based on InAs/InP QDs have been demonstrated [12,13] the properties of QDs need to be significantly improved in order to achieve superior device performances. Improvements, in particular, have to be made on the wetting layer charge leak, QD size uniformity, and shape control [14–16].

A promising alternative approach for the synthesis of QDs would be selective area growth (SAG), where the dots are formed in the mask openings, compared to when the formation process does rely on the lattice mismatch between the QD and matrix materials [17,18].

Diblock-copolymer lithography is a potentially scalable and very advantageous fabrication technique of uniform nanopatterns. Great effort has been and continues to be made with the aim to understand the principles of thin film polymer self-assembly and how to control its properties on demand [18–20]. The diblock-copolymer PS-b-PMMA (polystyrene-b-polymethyl methacrylate), which can form patterns down to 20 nm feature-sizes, including patterns of standing cylinders or spheres, was used for SAG of QDs [21,22]. Very recently, laser emission was demonstrated from a device based on this type of SAG QDs [23].

In this article, we report on the first realization of SAG QDs prepared with the assistance of a new diblock-copolymer, PS-b-PDMS (polystyrene-block-polydimethylsiloxane). Due to the significantly greater Flory-Huggins interaction parameter χ, it is possible to form patterns with even smaller characteristic sizes compared to the previously used PS-b-PMMA. Additionally, this copolymer has numerous other advantages, which would significantly simplify the BCP lithography process. Procedures have been developed allowing for direct deposition of the block copolymer film on the substrate, without any need for a brush layer to control the surface energy of the substrate; specific solvent vapor annealing promotes the self-assembly process and the pattern formation in the film [24,25]. Moreover, PDMS is transforming into a hard mask of silicon oxycarbide while PS cylinders are etched away by oxygen plasma, hence no deposition of additional dielectric layers is required [25,26]. Detailed optical characterization of resulting InAs/InP quantum dots grown by selective area droplet epitaxy allow evaluating dynamical carrier processes in the QD material in the 10-300K temperature range.

## 2. Experiment

The fabrication process consisted of two distinct stages: fabrication of a dielectric mask by diblock-copolymer lithography and epitaxial growth of QDs by metal-organic vapor phase epitaxy (MOVPE). All the samples were prepared on 2" n-type InP substrates with (001)-orientation. First, we epitaxially deposited 450 nm of InP. All the epitaxial processes were carried out using a low-pressure (60 Torr) Turbodisc® MOVPE system with trimethylindium (TMIn), arsine ($AsH_3$) and phosphine ($PH_3$) as precursors, and hydrogen ($H_2$) as carrier gas.

*Mask fabrication*: The mask fabrication process is schematically illustrated in Figure 1(a-c). PS-b-PDMS powder (61-111 K) was dissolved in cyclohexane 1/400 mg/ml and spin-casted onto the wafer forming an ultrathin film of 30 nm (Fig. 1(a)). Ellipsometry measurements verified the film thickness. To stimulate the polymer pattern formation, the film was placed into a chamber with methylcyclohexane saturated vapor until the pattern of standing cylinders is formed (Fig. 1(b)). Scanning electron microscopy (SEM) was applied to investigate the pattern.

To transform PS-b-PDMS patterned films into a hard mask, an Inductively Coupled Plasma-Reactive Ion Etching (ICP-RIE) process was carried out. A short step of sulfur hexafluoride plasma etch of 10 s was applied in the beginning to remove a thin PDMS skin layer on top of the polymer film and to open the PS cylinders [27]. The following oxygen plasma etch with a duration of 26 s removed the PS cylinders and oxidized the PDMS matrix. As a result, the PS-b-PDMS patterned film was transformed into the silicon oxycarbide hard mask with openings of 20-30 nm in diameter (Fig. 1(c)). The masks were afterwards inspected with a SEM and an atomic force microscope (AFM).

*Selective area epitaxy*: Prior to the selective epitaxial growth step, the wafer with the mask on top was cleaned by dipping it into concentrated sulfuric acid for 2 min with following rinsing in de-ionized water with nitrogen bubbles for 3 min. After drying with nitrogen gas, the wafer was loaded into the MOVPE chamber. Then, an annealing step was carried out under phosphine ambient at 550 - 650 °C for 15 min. The next step was in-situ InP etching by carbon tetrabromide ($CBr_4$) in the mask openings to form an array of holes (Fig. 1(d)). Prior the QD deposition the temperature was lowered to 450 °C and the sample was exposed to $AsH_3$ for 27 s. Indium droplets were deposited in the etched holes for 3 s with TMIn flow of 0.23 µmol/min in the absence of any $V^{th}$ group precursor in the growth chamber. Annealing indium droplets under $AsH_3$ for 60 s promotes their crystallization into InAs following the structure of the InP crystal lattice. InAs QDs were capped first by 3 nm, and next by 5 nm of InP deposited at 450 °C and 550 °C, respectively (Fig. 1(e)). After this, a short annealing at 550 °C in $PH_3$ ambient was done to promote an out diffusion of point defects formed because of the low growth temperature [28,29].

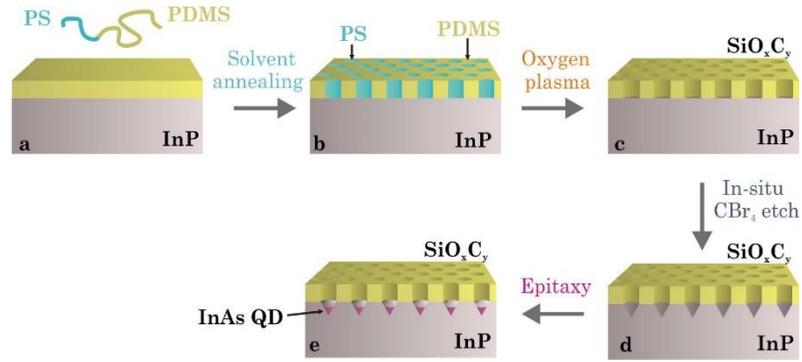

Fig. 1. Schematic illustration of the mask fabrication and QD growth process a-e.

*Optical characterization.* For the spectroscopic experiments, the structure with SAG InAs/InP QDs was held in a variable-temperature optical cryostat, with the temperature control within the range of 10-300 K. For time-integrated and time-resolved photoluminescence experiments, the sample was excited by a train of ~160 fs long pulses with ~13.2 ns pulse-to-pulse distance. The pulse train either came from a mode-locked Ti:Sapphire laser (Coherent Mira-HP) operating at the wavelength of 832 nm (~1.48 eV), or a synchronously pumped optical parametric oscillator (Coherent/APE OPO-HP), providing pulses with the wavelength of 1.4 μm (~0.89 eV). Emission from the structure was collected in a standard far-field optical setup and dispersed by monochromator with a focal length of 0.3 m (Princeton Instruments Acton SP2300i). The time-integrated photoluminescence (PL) spectra were measured with the lock-in technique (Signal Recovery 7560) at a carrier frequency of 2 kHz, with a liquid-nitrogen-cooled InSb photoconductive single channel detector (Hamamatsu P7751-02). Time-resolved photoluminescence (TRPL) is measured with a near-infrared streak camera system (Hamamatsu C112930-02) operating in the single-photon counting regime with a minimum time-resolution of ~20 ps.

## 3. Results and discussion

### 3.1. Fabrication

*Block-copolymer lithography.* The process of polymer self-assembly depends on the surface energy of the wafer. Contact angle measurements showed the difference in surface energy between epitaxy grown InP and $SiO_2$ surfaces. We discovered that the oxygen plasma treatment allowed to change the surface energy of InP to be similar to Si. In this way, the desired pattern of standing PS cylinders of 30 nm in diameter was achieved after annealing the PS-b-PDMS film with a thickness of 30 nm in methylcyclohexane saturated vapor for 55 min at 21 °C. The presence of short-range hexagonal order indicates a standing cylinder pattern, while the absence of such order is typical for the spherical polymer phase [30]. The self-assembly process of DBC is susceptible to the annealing time. Already a 5-minute deviation induces significant defects of a different polymer phase. Temperature and humidity also influence the self-assembly process, and the film thickness should obey the commensurability condition. A deviation in film thickness leads to terracing, which results in a nonuniform pattern phase [31].

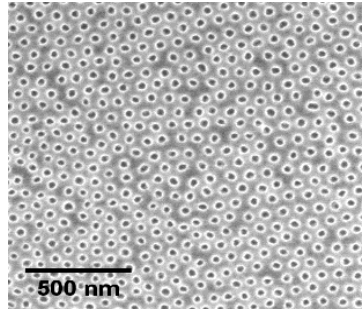

Fig. 2. SEM image of $SiO_xC_y$ hard mask.

After the pattern formation a top PDMS layer covers the pattern [32] and hence it is essential to remove this layer before the oxygen plasma step. Otherwise, the top layer will be oxidized and will not allow for etching of PS cylinders. To make sure that the mask openings are completely revealed after removing PS cylinders, the mask was investigated with an AFM equipped with ultrathin Bruker ScanAsyst-AIR tips.

*Growth of quantum dots.* The first step prior to an epitaxial growth is high-temperature annealing in phosphine overpressure to remove the native oxide from the semiconductor surface. However, annealing InP wafers covered by $SiO_xC_y$ mask at the standard temperature of 650 °C in phosphine ambient resulted in the formation of InP surface defects while the mask in those areas got membranized. A typical SEM image of such defects is shown in Fig. 3(a). By lowering the annealing temperature down to 550 °C the formation of those under-etch defects was avoided. A similar problem was described in [23], which in this case was attributed to $CBr_4$ under-etching, while in our case the high temperature is responsible for InP etching. During the annealing step, not only the native oxide but also InP in the mask openings is desorbed, and those thermo-etched areas from neighboring openings merge forming surface defects. Since the etch rate is different for different crystallographic planes the etched pits have faceting by opening (111) family planes. Those defects are clearly seen on the SEM image presented in Fig. 3(b), taken after the mask removal. Lower annealing temperatures can result in incomplete de-oxidation, but an increase in annealing time can compensate for this. The efficiency of de-oxidation was indirectly evaluated by measuring PL from the QD array. Comparison of the room temperature PL signal for identical QD samples grown after 650 °C annealing and 550 °C annealing indicates that the optical quality of the material is identical.

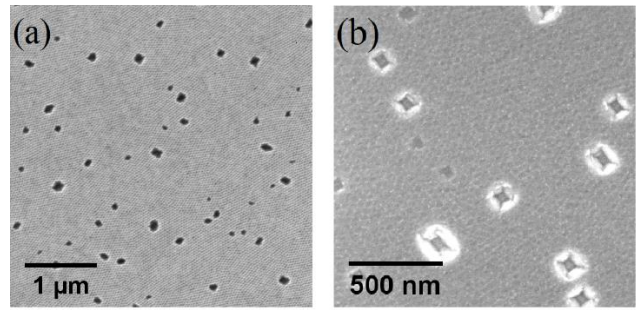

Fig. 3. (a) InP surface defects formed underneath of $SiO_xC_y$ mask after 650 °C annealing. (b) The same sample after the mask is removed.

For in-situ etching in the MOVPE chamber in the mask openings we employ $CBr_4$. There are two purposes for this. The first is to remove areas with a high density of point crystal defects after plasma etching and to improve the interface quality [33]. These point defects could act as nonradiative recombination centers. The second purpose is to etch holes in InP to bury and to control the shape of QDs to be grown in those holes. The etch process is self-limited and strongly depends on the crystallographic orientation, dominantly opening the (111) plane family [34]. The etching time of 20 s was kept short to avoid possible under-etching of InP under the mask and merging of etched pits [23].

### 3.2. Optical characterization

*Time-integrated photoluminescence studies.* Photoluminescence (PL) spectra obtained in the temperature range of 10-300 K for the investigated structure are presented in Fig. 4(a). The spectra are measured under conditions of the above InP barrier excitation and with a relatively high optical pumping power density ($P_{exc} \approx 5.5$ W/cm$^{-2}$). The PL spectrum from SAG InAs/InP

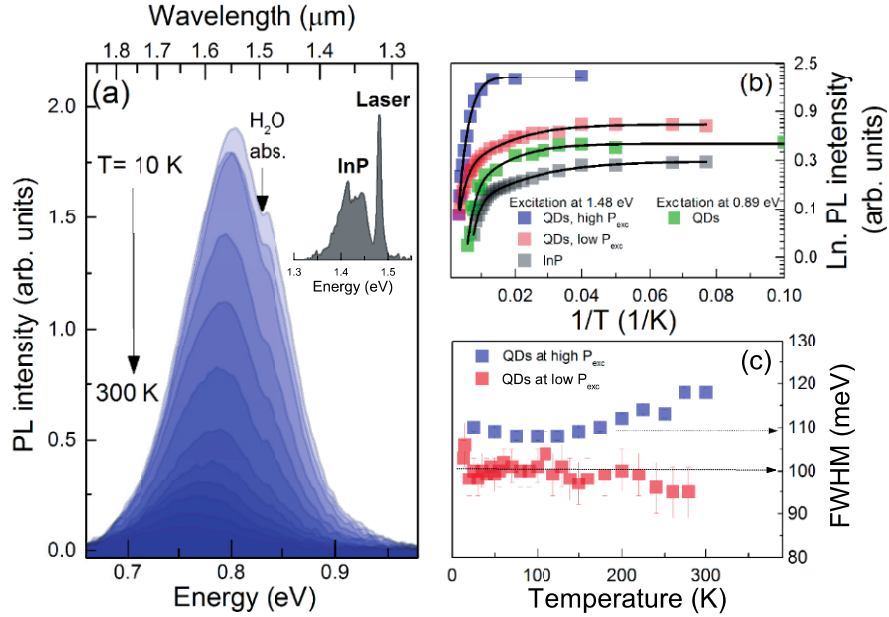

Fig. 4. (a) Temperature dependence of PL emission from SAG InAs/InP QDs ($E_{exc}$ = 1.48 eV, $P_{exc} \approx 5.5$ W/cm$^{-2}$). Inset: PL emission from the InP barrier and the laser spectrum at T = 10 K. (b) Temperature PL quench for QDs measured under high/low excitation power above InP barrier (red and blue squares) and below InP barrier excitation under low excitation power (green squares), and the PL quench for InP barrier under low excitation power (black squares). These dependences are arbitrary shifted on the intensity scale for better visibility. (c) Full-Width-at-Half-Maximum (FWHM) parameter for the QDs PL band under two optical pumping powers.

QDs is centered at ~0.8 eV (~0.76 eV) at T = 10 K (300 K) and covers the C- and L- telecom spectral windows. Each PL band was fitted to a Gaussian profile allowing for the extraction of the peak intensity (Fig. 4(b)) and its full width at half maximum - FWHM (Fig. 4(c)) in order to evaluate the carrier confinement in the dots.

Since the band offset of the strained InAs/InP material system is expected to be ~300 meV for the conduction band, and ~500 meV for the valence band [35], the carrier confinement in InAs/InP QDs should be rather strong. It is relevant to investigate the temperature impact on the carrier distribution of the QD ground state (GS) and the efficiency of the QD emission process. The PL intensity for the SAG QDs plotted as a function of inverse temperature, T, is shown in Figure 4 (b). Different regimes of excitation were investigated: (i) excitation energy tuned above the InP barrier, $E_{exc}$= 1.48 eV, and high excitation power density $P_{exc} \approx$ 5.5 W/cm$^2$ (blue squares), (ii) $E_{exc}$= 1.48 eV, and low $P_{exc} \approx$ 0.6 W/cm$^2$ (red squares), and (iii) quasi-resonant excitation of the dots, $E_{exc}$= 0.89 eV, nearly 0.53 eV below the InP bandgap and low $P_{exc} \approx$ 0.6 W/cm$^2$ (green squares). In addition, the PL thermal quench for InP is presented in Fig. 4(b) (black squares), obtained under $E_{ex}$ = 1.48 eV, and $P_{exc} \approx$ 0.6 W/cm$^2$. The PL quenches in Fig. 4(b) are arbitrarily shifted on the intensity scale for easier mutual comparison. Each experimental trend is fitted (solid black lines) by a widely-used steady-state solution for the thermal rate equation (Eq.1) expressed as:

$$I(T) = I_0 / \left( 1 + \sum_{i=1}^{n} C_i e^{E_{a,i}/k_B T} \right) \qquad (1)$$

where $C_i$ is a pre-exponential factor for the i$^{th}$ thermal activation process [36], $E_{a,i}$ is the activation energy related to the i$^{th}$ process, and $k_B$ is the Boltzmann constant. In all the cases, experimental data points and fitting curves well converged when two thermal activation processes are considered (i=1,2) in Eq. 1. The resulting best fitting parameters are summarized in Tab. I.

**Table 1. Parameters of the PL quench**

| PL emission | $E_{a,1}$, (meV) | $E_{a,2}$, (meV) | $C_1$ | $C_2$ | Excitation, (eV) | Pumping power |
|---|---|---|---|---|---|---|
| QDs | 38±9 | 109±50 | 25 | 582 | Above InP (1.48 eV) | High |
| QDs | 56±10 | 7±3 | 34 | 2 | Above InP (1.48 eV) | Low |
| InP | 54±10 | 6±1 | 34 | 2 | Above InP (1.48 eV) | Low |
| QDs | 61±23 | 9±3 | 23 | 3 | Quasi-resonantly in QDs (0.89 eV) | Low |

While the pre-exponential factors $C_i$ cannot be easily compared due to the different excitation regimes, one can compare the activation energies. For the PL thermal quench measured for SAG QDs under high and low $P_{exc}$ and $E_{exc}$= 1.48 eV, both $E_{a,1}$ activation energies can be considered as similar owing to the same activation mechanism that dominates the PL quench independently on the optical power density. In contrast, the $E_{a,2}$ energies are much different, indicating different activation processes. Their visibility in the overall PL quench depends on the occupation factor of the density of states (DOS) in the structure and the temperature range. On one side, the process with a higher activation energy $E_{a,2}$ = 109±50 meV can be tracked only under high photo-excitation and at T > 200 K. On the other side, the process with a low activation energy $E_{a,2}$= 7±3 meV is already present in the low-temperature regime 15 K< T < 100 K. Most likely, its visibility depends on the occupation factor of the DOS addressed during the photo-excitation. It is most intriguing to compare the PL quench parameters for the InP barrier and SAG QDs, since apparently, they are very similar within the uncertainty of the fitting parameters. This observation suggests that the recorded PL quench of SAG QD emission is controlled by thermal processes occurring in the barrier DOS. This conclusion is strengthened by comparison to the PL quench for QDs excited quasi-resonantly.

The respective activation energies are the same as those obtained for above-barrier excitation (see Tab. I). Therefore, the PL quench observed for SAG QDs cannot be discussed in terms of typical thermal activation mechanisms like carrier re-excitation from the QD GS to higher-lying states within the confinement potential of a dot or to the wetting layer. To describe the observed behavior, we suggest the following scenario: a population of photo-generated carriers is effectively trapped by an extensive DOS located within the InP bandgap, and subsequently, it is only partially transferred to QDs. With increasing temperature, a certain population of trapped carriers from their localized potentials releases in the barrier with the localization depth of roughly 7 and 60 meV (Tab. I). Thermally activated carriers, in their vast majority, are non-radiatively lost in the barrier and the efficiency of the processes increases with T. It causes that less and less carriers feed the QD states with increasing T. The nature of carrier localized states in the InP bandgap is related to various lithography steps like impurity levels and defect states at the native oxide/InP interface as well as structural defects formed in the InP matrix during ICP dry etch. Under the incomplete de-oxidation of InP, one can point to $In_P^{-1}$ [37] and $O_P^0$ that create trap states near the conduction band edge, where the former is localized ~60 meV below the conduction band edge of InP and the latter is located deep inside the InP bandgap, as suggested in [38]. The $In_P^{-1}$ defect level can be responsible for the observation of the ~60 meV activation energy causing the temperature-induced PL quench.

The temperature-induced quench investigation of SAG QD PL emission indicates that the barrier DOS has a significant impact on the optical characteristics of these dots and influences the carrier redistribution process among the QD ensemble. Figure 4 (c) presents the FWHM of the peak emission of SAG QDs. In the temperature range of 10-300 K, the FWHM changes only slightly within 5% of the value. From 10 K up to ~200 K the changes are even smaller, and the FWHM is considered to be constant for a given $P_{exc}$. Carrier redistribution processes among dots can be visible in the PL quench characteristics at T>200 K, and high $P_{exc}$. For $P_{exc} \approx$ 0.6 W/cm$^2$, the FWHM is decreasing with T, while for $P_{exc} \approx$ 5.5 W/cm$^2$, it increases. The former can be interpreted as a speedup of the carrier losses in the barrier owing to population of a smaller ensemble of QD states having a deeper confining potential. The latter can be related to population of a broader ensemble of QDs and their excited states, both in the vicinity of carrier migration in the barrier DOS.

*Time-resolved photoluminescence studies*. The results of the time-resolved (TR) PL experiments are summarized in Fig. 5. The excitation is provided at $E_{exc}$ = 1.48 eV, and $P_{exc} \approx$ 0.6 W/cm$^2$. The examples of intensity-normalized TRPL traces obtained at the center-of-mass of the PL emission from SAG QDs are presented in Fig. 5(a). Similar TRPL traces have been measured for different emission energies around the PL peak energy to evaluate the possible effect of carrier redistribution among QDs.

Each TRPL trace is characterized by a substantial delay after the photo-excitation during which the PL reaches its maximum amplitude, Fig. 5(b). The numerical fit to the experimental data, assuming a bi-exponential decay and single exponential rise, result in two decay times parameters: $\tau_{PLdec,1}$, $\tau_{PLdec,2}$, and the rise time parameter: $\tau_{rise}$, summarized in Fig. 5(c), and (d), respectively, plotted as a function of temperature and the emission energy.

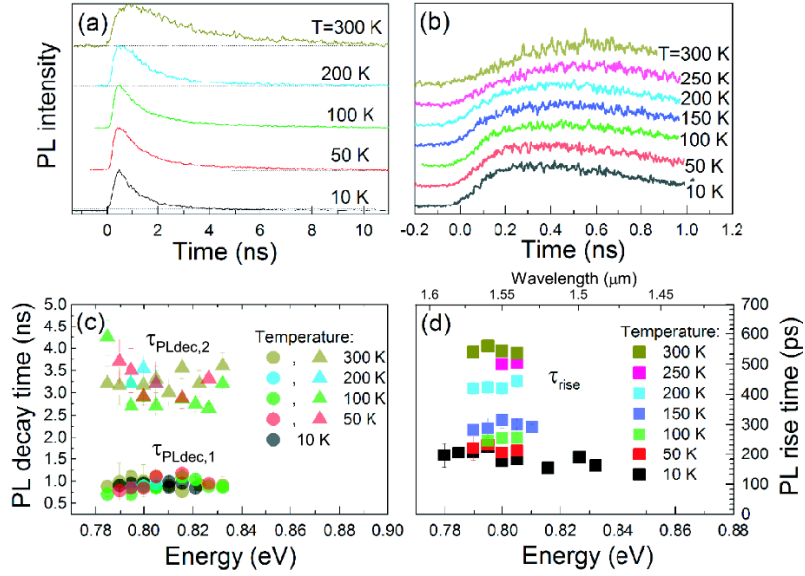

Fig. 5. (a) Time-resolved photoluminescence (TRPL) traces for selective-area growth InAs/InP QDs at various temperatures. (b) TRPL traces at their initial time-period after photo-excitation. (c) Dispersion of photoluminescence decay times at various temperatures, (d) dispersion of photoluminescence rise times as a function of temperature.

At T = 10 K, when the phonon population is negligible, one can assume that the PL dynamics are mostly determined by the confining potential of a QD. In this case, a TRPL trace is characterized by the PL rise time of ~200 ps, and a single PL decay time of ~1 ns. The second PL decay component cannot be resolved in this case due to its small amplitude. The PL rise time can be related to the sum of capture and the intra-band carrier relaxation times in the dot, after which the occupation of QD states reaches its quasi-equilibrium. Such an elongated $\tau_{rise}$ is quite unusual for QDs. For a self-assembled InAs/GaAs QD the intra-band relaxation time reaches several ps [39]. In a very recent study on intra-band carrier relaxation times in self-assembled InAs/InP QDs it was shown that relaxation times can be as short as 8 ps at cryogenic temperatures, even including the capture time [40]. For a comparable QD system like InAs/InAlAs/InP(001) [41] and InAs/InP(311B) [42], or InAs/InGaAsP/InP(001) quantum dashes [43], the intra-band relaxation time is reported within the range of 1-40 ps. However, in the SAG QDs, $\tau_{rise}$ can be dominated by the capture time, which further will be discussed by the temperature-induced trend presented in Fig. 5 (d). The PL decay is a more puzzling issue. On one hand, theoretical studies on an electron-hole (e-h) pair strongly confined in InAs/InP QDs have predicted the e-h radiative lifetime in the range of 1.6-2.4 ns depending on the dot size [44]. On the other hand, for large and strongly elongated InAs/InAlGaAs/InP QDs, where the e-h pair is in the intermediate confinement regime, the e-h can radiate into two superimposed lifetime components of ~1 ns and ~2 ns [45]. Although the latter would match with the obtained experimental results on SAG QDs, especially at higher temperatures, the estimated small size of the dots pushes the interpretation towards the e-h pair lifetime resulting from the strong confinement regime.

Figure 5(c) and (d) show temperature induced changes in the PL rise and decay times. One can clearly see that both components of the PL decay are barely changed with T, while the $\tau_{PLdec,1} \approx 1$ ns, the $\tau_{PLdec,2}$ is spread between 2.7 and 4.2 ns without any general trend. The spread is mostly generated by the accuracy of the fitting procedure. In general, each of the PL decay components is composed of radiative ($\tau_{rad}$) and non-radiative ($\tau_{non-rad}$) recombination processes:

$\tau_{PLdec}^{-1}=\tau_{rad}^{-1}+\tau_{non-rad}^{-1}(T)$. While the already discussed $\tau_{rad}$ is temperature independent, the $\tau_{non-rad}$ can strongly depend on T. Increasing the phonon population number with T leads to more efficient carrier-phonon scattering processes and thus shorter $\tau_{non-rad}$, resulting in a shorter $\tau_{PLdec}$. This typically leads to the observation of the time-integrated temperature-induced PL quench in self-assembled QDs, which is not the case for the studied SAG QDs where non-radiative recombination occurs predominantly in the barrier DOS. A similar effect has been observed for self-assembled InAs/GaAs QDs [46].

Another interesting point is the lack of any spectral dispersion for the obtained $\tau_{PLdec,1}(T)$, $\tau_{PLdec,2}(T)$, and $\tau_{rise}(T)$ functions presented in Fig. 5 (c) and (d). Along with the lack of a significant change in the FWHM parameter presented in Fig. 4 (c), it suggests negligible carrier migration between QDs and thus a relatively strong e-h confinement in the dot modified by the presence of the barrier DOS.

Finally, for the SAG QDs, the $\tau_{rise}$ is a strong function of T, as presented in Fig. 5 (d). With increasing T from 10 K to 300 K, the $\tau_{rise}$ increases from ~200 ps to ~560 ps. We previously argued that $\tau_{rise}$ is mainly dominated by the carrier capture time from the barrier to QDs. Additionally, we presented evidences for the existence of an extended DOS in the barrier that accumulate a significant portion of carriers after their photo-generation. These arguments lead to the conclusion that the elongated $\tau_{rise}$ is most likely related to charge migration within the barrier DOS before they are captured by QDs. A similar effect has been shown for self-assembled (In,Ga)As/GaAs QDs placed on a wetting layer with a large zero-dimensional density of states in there [47]. If the spatially distributed DOS outside the QDs is larger than the dots density, the carrier capture time can be controlled by a time-consuming temperature-induced hopping act among spatially distributed states in the barrier.

## 4. Conclusion

In conclusion, we present selective area growth of InAs/InP quantum dots assisted by block-copolymer lithography. The fabrication method of silicon oxycarbide hard masks with cylindrical openings of 20-30 nm in diameter on top of an InP substrate was developed. The pattern parameters could be regulated by manipulating the self-assembly process in PS-b-PDMS block-copolymer. We carried out a systematic investigation of the carrier dynamics in an array of SAG QDs by time-integrated PL and temperature dependent TRPL. The SAG QDs demonstrate strong carrier localization, however, their emission properties are affected by the defect-related density of states located in the InP barrier in the close vicinity of the dots. It is found that these states act as a carrier reservoir for the dots and dominate temperature-induced carrier migration in the structure. The effect is seen although of a strong PL quench from QDs accompanied by negligible changes in the PL decay time in the whole temperature range. This scenario is additionally supported by a slow and elongated temperature carrier capture time by the dots ranging from 200-500 ps. The nature of those defects is likely related to the technological process, the contamination with foreign atoms during lithography steps as well as damage of the crystal lattice during the ICP dry etching process.

## Funding

Villum Fonden. Young Investigator Programme. (VKR023442)

Polish budgetary funds for science in 2018-2020. "Diamond Grant" program (DI 2017 011747).

## Acknowledgements.

Authors thank Alexander Huck for fruitful discussions.